\documentclass[%
 reprint,
 superscriptaddress,
 amsmath,amssymb,
 aps,
 prl,
]{revtex4-2}

\usepackage{graphicx}
\usepackage{bm}
\usepackage{hyperref}

\newcommand{\Lplus}{\mathcal{L}_{+}}
\newcommand{\Lminus}{\mathcal{L}_{-}}
\newcommand{\Lboth}{\mathcal{L}_{\pm}}
\newcommand{\eul}{\mathrm{e}}

\begin{document}

\title{On the Doppler Effect in Static Materials:\\ Vortices and Black, White and Worm-Hole Like Wave Trajectories}

\author{Tristan Lawrie}
\email{t.lawrie@exeter.ac.uk}
\author{Simon A. R. Horsley}
\affiliation{Centre for Metamaterial Research and Innovation, Department of Physics and Astronomy, University of Exeter, United Kingdom}

\date{\today}

\begin{abstract}
We demonstrate the Doppler effect in a static material.
The effect is achieved with a network of thin waveguides whose lengths depend on the direction of travel, chosen to simulate the compression and expansion of the wave experienced by a source travelling with velocity $v$.
Each edge of length $\ell$ becomes a pair of waveguides of lengths $\ell/(1\mp v/c)$, one for each direction of travel.
On a lattice this gives the field pattern of a moving source.
To demonstrate the effect, we design a vortex, an analogue black hole and wormhole.
\end{abstract}

\maketitle

\section{Introduction}\label{sec:intro}

A source that moves through a medium radiates waves that are compressed ahead of it and stretched behind it.
Seen from the frame of the source the medium flows past, and waves travelling with and against the flow accumulate different phases over the same distance~\cite{Pierce2019,Leonhardt1999}.
Motion therefore makes a medium non-reciprocal~\cite{Caloz2018,Nassar2020}.
Since bulk motion at an appreciable fraction of the wave speed is rarely practical~\cite{Fleury2014}, active communities obtain the same physics from material properties that travel in space and time~\cite{Galiffi2022,Caloz2020,Engheta2021}, with examples ranging from non-reciprocal isolation~\cite{Sounas2017} and a synthetic Fresnel drag~\cite{Huidobro2019} to one-way propagation of elastic waves~\cite{Wang2018} and the eigenpulses of time-varying media~\cite{Horsley2023}.

In this Letter we propose a medium that reproduces the propagation of waves in a moving frame while remaining completely static.
We propose the static material as a network of thin waveguides, which we model via the scattering language of quantum graph theory~\cite{Kottos1997,Gnutzmann2006,Berkolaiko2013}, a model that has proved reliable in the study of metamaterials~\cite{Lawrie2022,LawrieThesis}, in both acoustic~\cite{Lawrie2024} and microwave~\cite{Hul2004,Lawniczak2026} networks.
The theory supplies only the mathematics: the waves are classical, and the network stands as a model of a continuous medium, the lengths of its edges standing in for material parameters such as $\varepsilon$ and $\mu$.

The network can be viewed as a graph, whose vertices act as junctions joining its edges, and whose edges act as waveguides [Fig.~\ref{fig:chain}(a)].
On a graph the only property of an edge that modifies wave propagation is the accumulated phase.
We show that the phases of a moving frame are obtained from two ingredients: a way of sending the two directions of travel along an edge on separate paths, and a difference in the lengths of those paths [Fig.~\ref{fig:chain}(b)].
Nothing restricts the two lengths to those of a uniform motion [Fig.~\ref{fig:chain}(c)].
We therefore obtain the Doppler effect in a medium in which nothing moves, and demonstrate the reach of the model with four examples: the Doppler effect on a periodic lattice, a vortex from which waves leave a source as a spiral, an analogue black hole, and a wormhole.
Because the velocity can be chosen edge by edge, a network can emulate flows that would be hard to set up in any real medium, which opens the physics of moving media and analogue gravity to anyone able to build a network of waveguides, for sound, microwaves or light, without modulating anything in time.

This Letter is structured as follows.
Section~\ref{sec:lengths} shows how the shift in frequency of the Doppler effect can be reproduced by a change of length, and Sec.~\ref{sec:networks} demonstrates the Doppler effect on a two-dimensional lattice.
Section~\ref{sec:applications} extends the method to a vortex, an analogue black hole and a wormhole, and Sec.~\ref{sec:conclusion} concludes.

\section{Effective motion from variable lengths}\label{sec:lengths}

We begin with the one-dimensional wave equation, which describes waves throughout classical physics: sound in ducts and pipes~\cite{Pierce2019}, voltages and currents on transmission lines and electromagnetic waves in waveguides~\cite{Pozar2012}, longitudinal waves in elastic rods, transverse waves on strings and long water waves in shallow channels.
In a medium at rest it reads
\begin{equation}
\partial_t^2\phi-c^2\partial_x^2\phi=0 ,
\label{eq:wave}
\end{equation}
where $c$ is the wave speed.
What follows is therefore not tied to any one kind of wave.
Its solution is a plane wave $\phi=\eul^{i(kx-\omega t)}$ with $k=\omega/c$.
Suppose now that a source moves through the medium with velocity $v$, a loudspeaker carried through still air, say.
In the frame of the source the medium flows past, so the wave speed relative to the source is $c-v$ ahead of it and $c+v$ behind it.
Assuming non-relativistic velocities, the Galilean transformation $x\to x+vt$ to this frame turns Eq.~(\ref{eq:wave}) into
\begin{equation}
\left(\frac{\partial}{\partial t}-v\frac{\partial}{\partial x}\right)^{2}\phi-c^{2}\frac{\partial^{2}\phi}{\partial x^{2}}=0 ,
\label{eq:moving}
\end{equation}
whose solutions are $\phi=\eul^{i(\kappa_\pm x-\omega t)}$, for $|v|<c$, with
\begin{equation}
\kappa_{\pm}=\pm\frac{k}{1\mp M},\qquad M=\frac{v}{c},
\label{eq:kappa}
\end{equation}
for waves travelling ahead of ($+$) and behind ($-$) the source, where $M$ is the Mach number of the source.

Across a lattice spacing of length $\ell$ the two waves accumulate the phases $|\kappa_\pm|\ell=[\omega/(1\mp M)]\,\ell/c$.
Since the frequency and the length are multiplied in the exponent, the effective change of frequency produced by the motion can be written instead as an effective change of length,
\begin{equation}
\eul^{i|\kappa_\pm|\ell}\rightarrow\eul^{ik\Lboth},\qquad \Lboth=\frac{\ell}{1\mp M},
\label{eq:L}
\end{equation}
which moves the velocity out of the wavenumber and into the length: a wave that travels a distance $\Lplus$ in one direction and $\Lminus$ in the other, in a medium at rest, acquires exactly the phases of the moving frame.
Since $\eul^{i\omega\ell/(c\mp v)}=\eul^{i\omega\Lboth/c}$ at every frequency, lines whose wave speed $c$ does not depend on frequency, such as the TEM mode of a coaxial cable~\cite{Pozar2012} or the plane-wave mode of an acoustic duct~\cite{Pierce2019}, reproduce the two travel times as well as the phases.

\begin{figure*}
\includegraphics[width=\linewidth]{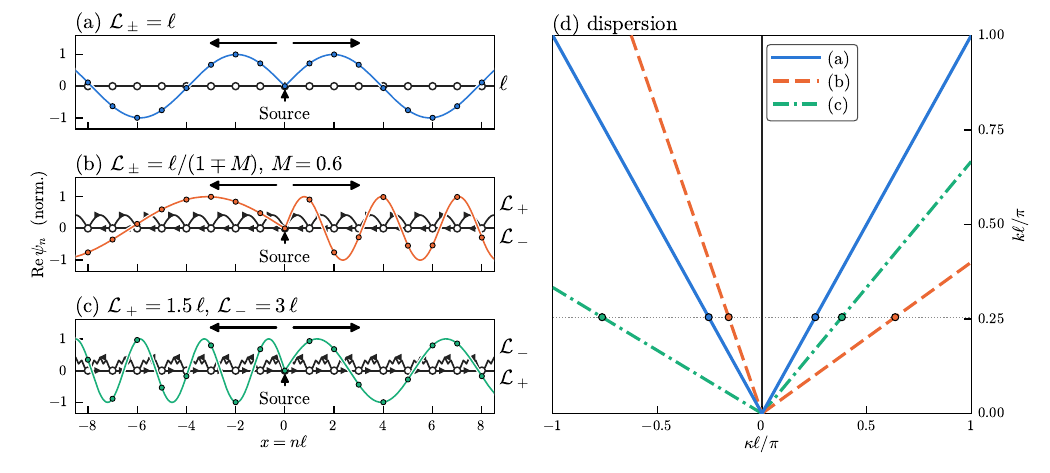}
\caption{The chain at $k\ell=0.8$. (a)--(c) The chain in black, with its edge lengths above each panel, and the real part of the field in colour; the dots are the solutions at the vertices, joined by an interpolation. (d) The dispersion of the three chains; the dotted line marks $k\ell=0.8$.}
\label{fig:chain}
\end{figure*}

The simplest network is a chain of vertices a distance $\ell$ apart [Fig.~\ref{fig:chain}(a)].
At each vertex the wave is continuous and the outward derivatives of the field sum to zero~\cite{Kottos1997,Lawrie2022}; with two edges per vertex this lets a wave pass without reflection, so a Bloch wave of wavenumber $\kappa$ gains the phase $\kappa\ell$ per cell and $\kappa=\pm k$, the dispersion of Eq.~(\ref{eq:wave}) [Fig.~\ref{fig:chain}(d)].
If instead each edge is split into two one-way lines~\cite{Tanner2001}, of lengths $\Lplus$ and $\Lminus$, a wave travelling to the right gains $k\Lplus$ per cell and one travelling to the left $k\Lminus$.
The Bloch wavenumbers of the chain are then exactly the $\kappa_\pm$ of Eq.~(\ref{eq:kappa}), up to multiples of $2\pi/\ell$, which is why we keep the symbol $\kappa$ for the Bloch wavenumber below, and the dispersion tilts over [Figs.~\ref{fig:chain}(b) and \ref{fig:chain}(d)].
This is the dispersion of Eq.~(\ref{eq:moving}), reproduced by a network in which nothing moves: a point source on the chain radiates waves that are compressed ahead of it and stretched behind it.
Nothing restricts the two lengths to the pair of Eq.~(\ref{eq:L}): any other choice tilts and rescales the two branches independently [Figs.~\ref{fig:chain}(c) and \ref{fig:chain}(d)], and on a lattice the lengths can be chosen afresh on every edge.

It is worth stating precisely what the network reproduces.
A static network is invariant in time, so a source of frequency $\omega$ drives a field of frequency $\omega$ everywhere, and a detector at rest in the network records no new frequency.
The network reproduces instead the moving source in its own frame, which is monochromatic and in which the Doppler effect appears as the two unequal wavenumbers $\kappa_\pm$ of Eq.~(\ref{eq:kappa}).
The familiar shifted frequencies $\omega/(1\mp M)$, heard by an observer at rest in the medium, follow on transforming back to the frame of the medium.
What is traded for a difference in length is this inequality of wavenumbers.

Direction-dependent lengths require the two directions of travel along an edge to be routed onto different paths, that is, time-reversal symmetry to be broken on every edge.
This is a general requirement that is met in many wave domains: by magnetically biased materials~\cite{Pozar2012}, by transistor-loaded structures under a constant bias~\cite{Kodera2011}, by coupled ring resonators for light~\cite{Hafezi2011}, and, for sound and water waves, by a flow, the natural counterpart of the motion considered here, which has been used to great effect in analogue-gravity experiments~\cite{Weinfurtner2011,Torres2017} and in acoustic circulators~\cite{Fleury2014}.
Should the reader want a concrete picture, a natural one is the three-port circulator [Fig.~\ref{fig:lattice}(e)]: placed at each end of an edge, it sends what a vertex emits onto one line and passes what arrives on the other line back to the vertex.
Circulators are standard microwave components~\cite{Pozar2012}, already used in networks of coaxial cables to break time-reversal invariance~\cite{Rehemanjiang2016,Bialous2016}, networks in which the scattering formalism used here has been confirmed experimentally~\cite{Lawniczak2026}.
The circulator is only an example, however: nothing in what follows is specific to cables or to circulators.
The one-way lines can also be made without a magnetic bias by modulating a waveguide in space and time~\cite{Yu2009,Sounas2017}, as in magnet-free circulators~\cite{Estep2014} and one-way elastic waveguides~\cite{Trainiti2016,Nassar2017,Wang2018}; the modulation is then confined to the elements that separate the two directions, while the network itself stays static.

\section{Waves in networks}\label{sec:networks}

\begin{figure*}
\includegraphics[width=\linewidth]{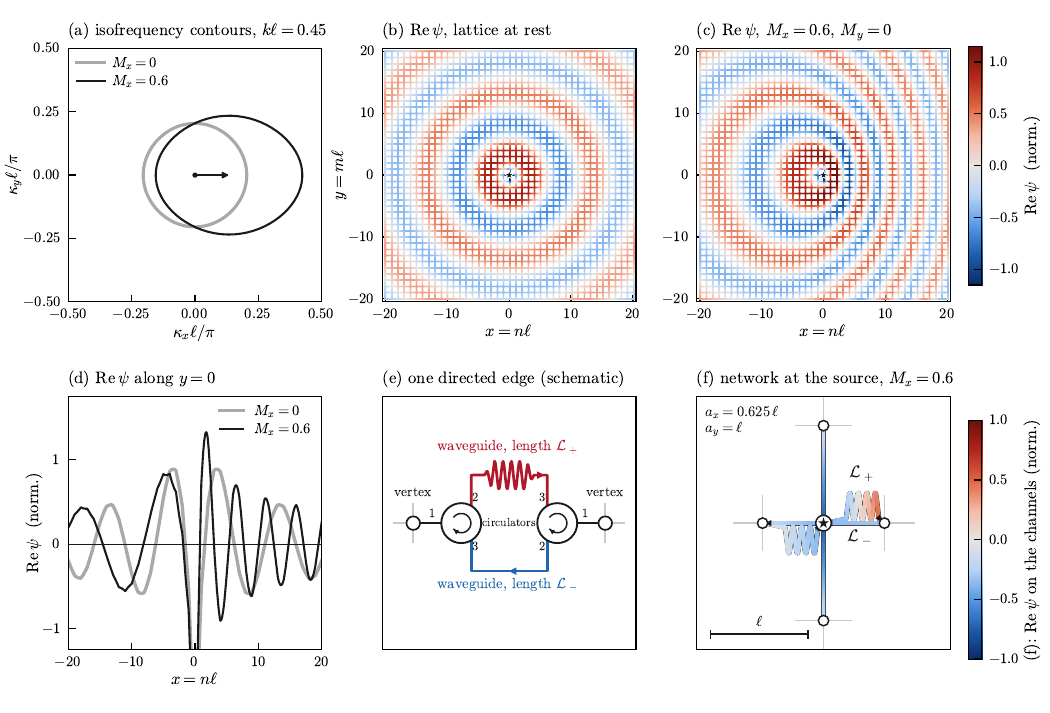}
\caption{A point source on a square lattice at rest and with $M_x=0.6$, $M_y=0$. (e) One edge built from two circulators. (f) The edges around the source, drawn to scale.}
\label{fig:lattice}
\end{figure*}

We consider a two-dimensional square network (Fig.~\ref{fig:lattice}) whose vertices (waveguide junctions) are joined by edges (thin waveguides) of direction-dependent lengths $\Lplus>\ell>\Lminus$; the longer lines act as space-coiled waveguides~\cite{Liang2012,Xie2013} [Fig.~\ref{fig:lattice}(f)].
A vertex is indexed by $n$ in the horizontal direction and $m$ in the vertical direction, and sits at the standard Cartesian coordinates $x=n\ell$ and $y=m\ell$ of the medium that the network emulates.
At every vertex the wave is continuous, $\phi_e(0)=\phi_{e'}(0)$ for connecting edges $e,e'$, and flux is conserved, the outward derivatives of the wave fields summing to zero, $\sum_e\phi_e'(0)=0$~\cite{Kottos1997,Lawrie2022}.
The field of a point source is the Green's function of the graph~\cite{Lawrie2023,LawrieThesis}, which we compute by solving the scattering equations of the network~\cite{Lawrie2022,LawrieThesis}.

\subsection{Lattice dispersion}\label{sec:dispersion}

Suppose the lattice is infinite and periodic, with period $\ell$ in both $x$ and $y$.
A Bloch wave then takes the value $\eul^{i(\kappa_xn+\kappa_ym)\ell}$ at vertex $(n,m)$, relative to its value at the origin.
Applying the vertex conditions at one vertex, with the far end of each edge related to its near end by this Bloch phase, gives the dispersion relation~\cite{SM,Lawrie2026}
\begin{equation}
\sum_{j=x,y}\frac{\cos\vartheta_j-\cos(\kappa_j\ell-\Delta_j)}{\sin\vartheta_j}=0 ,
\label{eq:D}
\end{equation}
where
\begin{equation}
\vartheta_j=\frac{k\ell}{1-M_j^{2}},\qquad \Delta_j=M_j\vartheta_j ,
\label{eq:theta}
\end{equation}
are, respectively, the mean and the half-difference of the two phases $k\Lboth$ of the edges along direction $j$.
The lengths are set by the velocity $\mathbf{v}=(v_x,v_y)$, of speed $v=\sqrt{v_x^2+v_y^2}$: the edges along $j$ take the lengths of Eq.~(\ref{eq:L}) with $M_j=v_j/c$.
At rest $\Delta_j=0$ and Eq.~(\ref{eq:D}) is the known dispersion relation of the square graph~\cite{Lawrie2022}.

Figures~\ref{fig:lattice}(b) and \ref{fig:lattice}(c) show the field of a point source on the lattice at rest and with $M_x=0.6$ and $M_y=0$.
With the velocity applied the isofrequency contour is displaced by $\Delta_x$ [Fig.~\ref{fig:lattice}(a)], and the wavefronts bunch ahead of the source and spread out behind it [Figs.~\ref{fig:lattice}(c) and \ref{fig:lattice}(d)].
This is the pattern of a source in motion, radiated by a network in which nothing moves.

\section{Applications of the Doppler lattice}\label{sec:applications}

Having shown that the Doppler effect can be achieved in a static network by a choice of waveguide lengths, we now use the freedom to choose a different velocity on every edge.
The examples below are flows that are hard to set up in a real medium, let alone to modulate in time: a vortex, in which the velocity circulates about a point, an inward flow that cannot be escaped inside a horizon, and a wormhole made by joining such a horizon to its time reverse.
Such flows are studied in analogue gravity~\cite{Unruh1981,Weinfurtner2011,Torres2017} and in the design of non-reciprocal devices~\cite{Caloz2018,Nassar2020}, and all are built from the networks of Fig.~\ref{fig:lattice}.

\subsection{A vortex}\label{sec:vortex}

\begin{figure*}
\includegraphics[width=\linewidth]{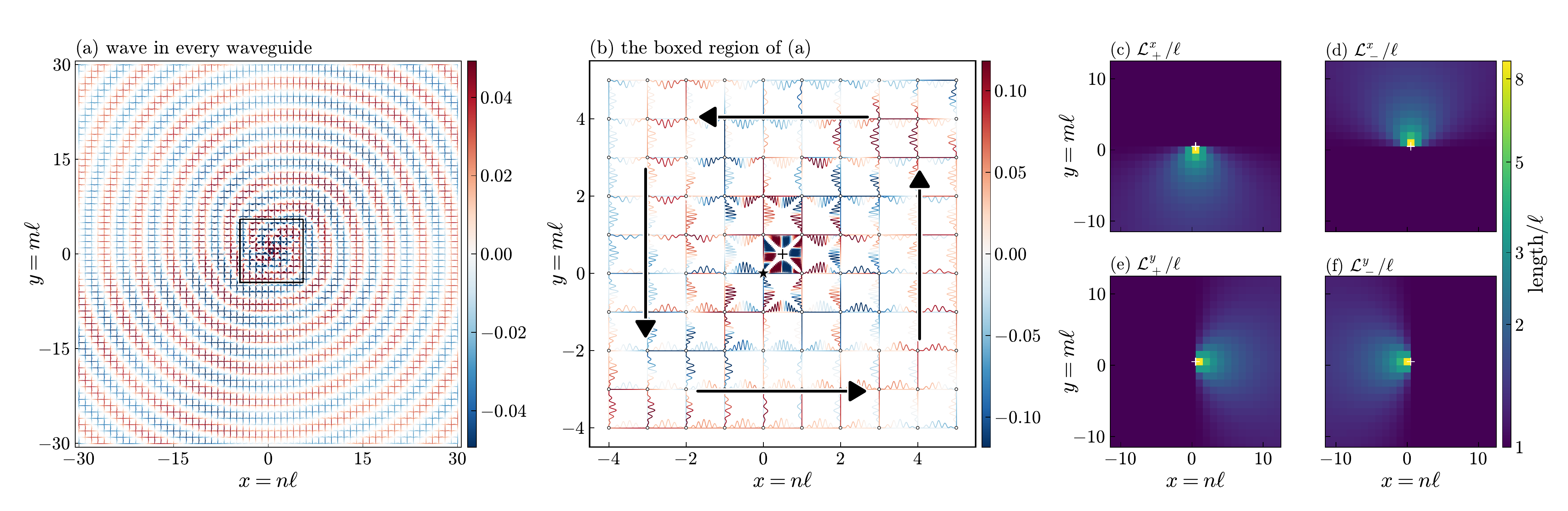}
\caption{The vortex of Eq.~(\ref{eq:vortex}) with $q=2$, at $k\ell=0.8$, driven by a point source at $n=m=0$; the cross marks the core. Each waveguide is drawn with its true length, the two of each edge on either side of the line joining its vertices; (c)--(f) have a logarithmic colour scale.}
\label{fig:vortex}
\end{figure*}

We keep the vertices of a square lattice at the spacing $\ell$ and let the longer lines circulate around the centre of one cell, the core [Figs.~\ref{fig:vortex}(a) and \ref{fig:vortex}(b)].
If an edge subtends the angle $\delta\theta$ at the core, its shorter line has length $\ell$ and its longer line has length
\begin{equation}
\mathcal{L}_{\rm long}=\ell+\frac{|q|\lambda_0}{\pi}\,|\delta\theta| ,
\label{eq:vortex}
\end{equation}
where $\lambda_0$ is the wavelength at a design frequency $\omega_0$ and $q$ is an integer whose sign selects the sense (the direction of circulation, clockwise or anticlockwise) in which the longer lines run round the core: anticlockwise for $q>0$ and clockwise for $q<0$ [arrows in Fig.~\ref{fig:vortex}(b)].
Far from the core an edge at the distance $r\gg\ell$, whose direction makes the angle $\alpha$ with the radius, subtends $\delta\theta\approx\ell|\sin\alpha|/r$, so that the extra length falls off as $1/r$ [Figs.~\ref{fig:vortex}(c)--\ref{fig:vortex}(f)].
Round any loop that encloses the core the two senses then differ in path by $2|q|$ wavelengths, and round any other loop they do not differ at all.
At the design frequency the field is therefore
\begin{equation}
\psi_{n,m}=\eul^{iq\theta_{n,m}}\,\tilde\psi_{n,m},
\label{eq:spiral}
\end{equation}
up to a constant phase, where $\theta_{n,m}$ is the angle of the vertex about the core and $\tilde\psi$ is the field of the same source in the reciprocal network in which both lines of every edge have their mean length~\cite{SM}.
The phase of the wave winds $q$ times round the core, and the wavefronts of a point source become a spiral with $|q|$ arms [Fig.~\ref{fig:vortex}(a), for $q=2$].

The spiral is a pattern of phase: the intensity on every line equals that of the reciprocal network, so that no additional power circulates.
Since the lengths are fixed, the accumulated phase scales as $\omega/\omega_0$, and Eq.~(\ref{eq:spiral}), with $q$ replaced by $q\,\omega/\omega_0$, holds at the frequencies for which $q\,\omega/\omega_0$ is an integer.
The winding therefore grows with frequency, but at every such frequency it has the sign of $q$: the sense of rotation is fixed when the lines are cut.

\subsection{An analogue black hole}\label{sec:blackhole}

In the acoustic analogue of a black hole~\cite{Unruh1981} the medium flows inward, a wave that travels outward is slowed by the flow, and where the flow reaches the wave speed, at the horizon, it no longer escapes.
In Eq.~(\ref{eq:L}) this is the limit $M\to1$, in which the line that carries waves against the flow becomes infinitely long, $\Lplus=\ell/(1-M)\to\infty$.

\begin{figure*}
\includegraphics[width=\linewidth]{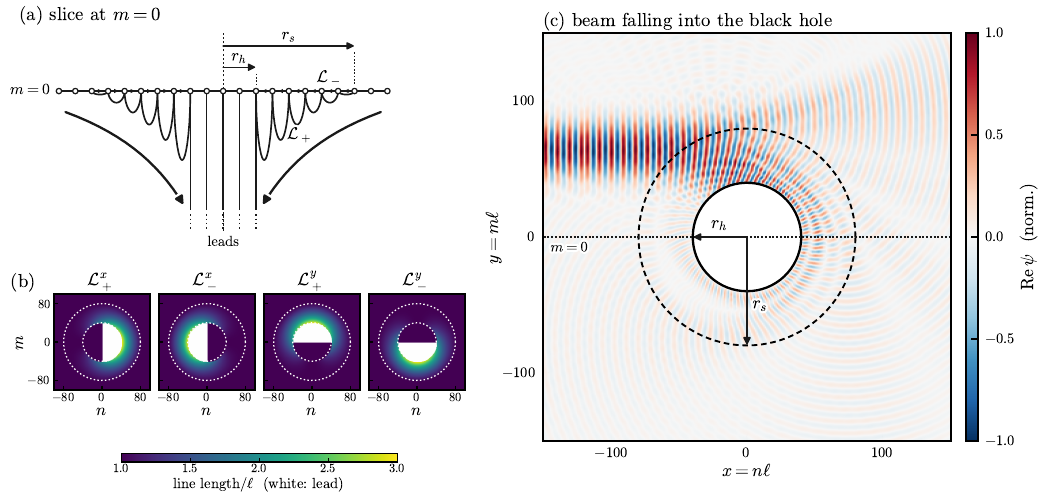}
\caption{The analogue black hole at $k\ell=0.5$. (a) A slice, not to scale, with the longer line of each edge hanging below the plane. (c) Real part of the wave for a Gaussian beam of half-width $20\,\ell$, aimed $64\,\ell$ from the centre.}
\label{fig:blackhole}
\end{figure*}

On the graph we place the black hole at $x=y=0$ and define the edge lengths by the radius $r=\sqrt{x^2+y^2}$ [Fig.~\ref{fig:blackhole}(a)].
The radius of the black hole is $r_h=40\,\ell$.
Within it, $r<r_h$, the outward-directed lines are infinite in length, creating leads (semi-infinite edges) down which a wave leaves and from which it cannot return, in practice matched loads that absorb it [Fig.~\ref{fig:blackhole}(a)], while the inward-directed lines keep the length $\ell$.
The radius of the shell is $r_s=80\,\ell$.
Within the shell, $r_h\le r<r_s$, we introduce the effect of the black hole as lengths that increase as we move closer to $r=0$.
We do not attempt to reproduce a particular spacetime, such as the Schwarzschild metric: the lengths are simply a choice that attracts waves one way and not the other.
They follow from a continuous profile discretised onto the edges, the mean length of an edge set by its midpoint and the difference of its two lengths by its two ends~\cite{SM}, which to leading order gives
\begin{equation}
\mathcal{L}^j_\pm=\ell\left[\frac{r_s}{r}\pm\left(\frac{r_s}{r}-1\right)\frac{j}{r}\right],
\label{eq:hole}
\end{equation}
where $j=x$ or $y$, and $\mathcal{L}^j_\pm$ are the lengths of the lines of an edge along $j$ that carry waves towards $\pm j$.
An edge that points at the black hole therefore has an inward line of length $\ell$ and an outward line of length $(2r_s/r-1)\,\ell$, so that a wave travels in freely but is held back on its way out, while an edge at right angles to the radius has two equal lines of length $\ell r_s/r$, which bend a passing wave inward.
Every line outside the black hole is finite, the longest being the outward lines at $r=r_h$, of length $(2r_s/r_h-1)\,\ell=3\,\ell$.
Beyond the shell, $r\ge r_s$, every line has the length $\ell$ [Fig.~\ref{fig:blackhole}(b)].
The coordinates $x=n\ell$, $y=m\ell$ and the time are those of the laboratory, in which the black hole is at rest and the medium flows through it.

Figure~\ref{fig:blackhole}(c) shows a beam sent past the structure on a lattice of $381\times381$ vertices.
The beam is aimed well outside the black hole, yet it bends inward as the lines lengthen and spirals in.
The structure behaves as a black hole for the waves it carries: a wave falls in easily but cannot get out, and a source placed inside $r_h$ produces no field outside it, which no reciprocal structure, such as the graded-index optical black hole of Narimanov and Kildishev~\cite{Narimanov2009}, can do.

What it does not produce is Hawking radiation~\cite{Hawking1975}.
In a flowing medium this radiation is born at the horizon, where a wave is converted into a pair: a wave of positive frequency in the frame of the medium, which escapes, and a partner of negative frequency, which is swept inside~\cite{Unruh1981,Weinfurtner2011}.
Since the network trades a change of frequency for a change of length, beyond the horizon, $M>1$, Eq.~(\ref{eq:L}) requires negative lengths, $\ell/(1-M)<0$, on which a wave spends a negative time, and it is on these lines that the partner would travel.
Negative lengths are therefore the requirement for Hawking radiation in the network.
In the configuration chosen here they do not exist, the region beyond the horizon being replaced by leads, so the black hole captures waves as a horizon does but does not radiate; negative lengths would require elements that are active or modulated in time.

\subsection{A wormhole}\label{sec:wormhole}

\begin{figure*}
\includegraphics[width=\linewidth]{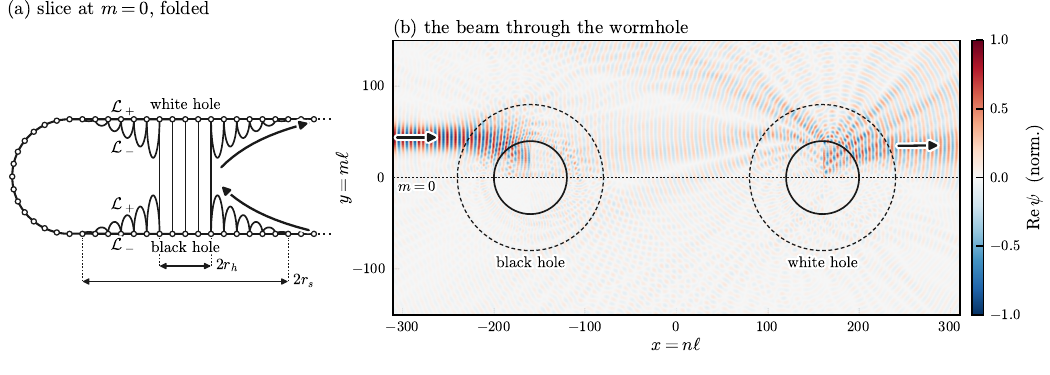}
\caption{The wormhole at $k\ell=0.5$. (a) The slice at $m=0$, folded and not to scale; the vertical lines are the throat. (b) Real part of the wave for a Gaussian beam of half-width $14\,\ell$, aimed $44\,\ell$ from the line $m=0$; the right-hand arrow marks the direction in which the wave that came through the throat leaves the white hole.}
\label{fig:wormhole}
\end{figure*}

The leads inside the black hole carry waves out of the network, and they need not end in absorbers.
In the same network, a distance $320\,\ell$ from the black hole, we place its time reverse: the same structure with the two lengths of every edge exchanged, $\Lplus\leftrightarrow\Lminus$.
This is a white hole: in its shell the outward lines are the short ones and inside $r_h$ the inward lines are the infinite ones, so that waves can only leave it.
Joining each lead of the black hole, by lines of length $\ell$, to the lead of the white hole at its mirror image in the line midway between the two holes turns the leads into one-way lines that connect the two holes, the throat of a wormhole.
Drawn with the network folded so that the white hole lies above the black hole, each lead joined to the one directly above it, this is the familiar picture of a wormhole [Fig.~\ref{fig:wormhole}(a)]; the fold is only a drawing, the vertices keeping the spacing $\ell$ around it, so that there is no curvature between the two holes, the lattice between them being the plain lattice.

In Fig.~\ref{fig:wormhole}(b) the beam falls into the black hole and re-emerges from the white hole.

\section{Conclusion}\label{sec:conclusion}

We have shown that the propagation laws of a moving medium can be written into a static network.
The construction is a single rule, Eq.~(\ref{eq:L}), which trades a velocity for a difference between the lengths of the two directions of travel: a network of waveguides whose lengths depend on the direction of travel propagates waves as a moving medium does.
Because the rule acts edge by edge, any velocity field can be written into a network, and the existing methods for designing networks, from Green's functions~\cite{Lawrie2023,LawrieThesis} to inverse design~\cite{Lawrie2025,Lawrie2026}, apply to it unchanged.
On a chain and on a periodic lattice the rule gives the dispersion and the field pattern of a moving source; when the lengths vary in space it gives a spiral wave of prescribed winding, a structure that draws in a passing beam and returns nothing, and a wormhole through which the beam passes from the black hole to a white hole.
None of these requires anything to move or to be modulated.

Since only the phase accumulated along each edge enters, the results hold for any classical wave guided by a network whose two directions of travel can be given separate paths, whether by a magnetic bias, by active elements or by a flow.
Because a velocity can be assigned edge by edge, static networks offer a simple laboratory for waves in flowing and curved media.

\section*{Acknowledgements}

Any data and code generated for this paper can be made available upon reasonable request. The authors acknowledge the financial support of the META4D Program Grant (EP/Y015673/1).

\end{document}